# *Solving single molecules: filtering noisy discrete data made of photons and other type of observables*

*Ophir Flomenbom, Flomenbom-BPS Ltd, 19 Louis Marshal St., Tel Aviv, Israel 66268*

**Abstract** In numerous systems in biophysics and related fields, scientists measure (with very smart methods) individual molecules (e.g. biopolymers (proteins, DNA, RNA, etc), nano – crystals, ion channels), aiming at finding a model from the data. But the noise is not solved accurately in not so few cases and this may lead to misleading models. Here, we solve the noise. We consider two state photon trajectories from any *on off* kinetic scheme (KS): the process emitting photons with a rate $\gamma_{on}$ when it is in the *on* state, and emitting with a rate $\gamma_{off}$ when it is in the *off* state. We develop a filter that removes the noise resulting in clean data also in cases where binning fails. The filter is a numerical algorithm with various new statistical treatments. It is based on a new general likelihood function developed here, with observable dependent form. The filter can solve the noise, in the detectable region of the rate space: that is, we also find a region where the data is "too" noisy. Consistency tests will find the region's type from the data. If the data is ruled "too noisy", binning obviously fails, and one should apply simpler methods on the raw data and realizing that the extracted information is partial. We show that not applying the filter while cleaning results in erroneous rates. This filter (with minor adjustments) can solve the noise in any discrete state trajectories, yet extensions are needed in "tackling" the noise from other data, e.g. continuous data and FRET data.

The filter developed here is complementary with our previous projects in this field, where we have solved clean two state data with the development of reduced dimensions forms (RDFs): unique models that are canonical forms of two state data, and with the development of a statistical and numerical toolbox that builds a RDF from *finite*, clean, two - state data. Thus, only the combined procedures enabling building the most accurate model from noisy trajectories from single molecules

**I. Introduction.-**

**I.1. The system.-** Nowadays, scientists measure smartly many processes in biology, chemistry and physics at a level of individual molecules, enabling (in principle) extracting information that was not accessible in the past about microscopic processes. Yet, finding the model from the signal is complicated. If the noise is not solved accurately in data from individual molecules, misleading models are reported. In this paper we solve ("tackle") the noise in discrete $m$ (=2,3,...) state data.

When talking about relevant experiments and processes, we list: * the passage of ions and biopolymers through individual channels [3 - 6], * activity and conformational changes of biopolymers (including fluorescence resonance energy transfer (FRET), atomic force microscopy (AFM) and other techniques) [1 - 2, 7 - 21, 59], * diffusion of molecules [22 - 25], * blinking of nano-crystals [26-29], etc.

Signals are time trajectories made of several discrete values, or states, where a popular example is of *on-off* trajectories (this is also the simplest example): trajectories that are made of *on* and *off* periods (also termed residence times or jumping times (JT); see Fig. 1A-1B). Extensions include FRET trajectories (that can have 3 and more states) and even continuous trajectories (where the trajectory is a continuous coordinate). From a noisy trajectory, one aims at finding the mechanism that can generate the observed process, has a physical sense and can supply scientific insights on the observed process. In many cases, we say that the model of the observed process is a multi − substate, multi - state Markovian kinetic scheme (KS). *On − off* KSs are popular [30-45], see Fig. 1E. The KS can stand, for example, for one of the following physical realizations: a discrete conformational energy landscape of a biological molecule, steps in a chemical reaction with conformational changes or environmental changes, quantum states, etc.

The aim is therefore building the mechanism from the noisy data. In many projects, we can solve the system only when solving data from individual entities, since other (experimental or numerical) techniques are missing or are not so informative. The case of enzymes is such an example. If the noise in the data is not filtered accurately when analyzing the data (due to ignoring the noise, or assuming that the noise is small, or using too simple filters), misleading models, partial and questionable conclusions are reported. Let me present an example: in [11], the authors did not filter the noise and used a binning method when building from the data the observable. They started with the raw data (photon arrival times), binned every following 100 photon arrival times, calculated the average, and used a specific model when building (from this quantity) a time dependent distance among residues in a protein. We can ask: What are the results when we bin 16 photons, 39 photons, 333 photons, in the computations of the rate?? What about filtering background photons?? What are the model independent quantities that we can extract from the data?? All these crucial points are not addressed in that project [11]. The simple binning method used in [11] might lead to misleading conclusions. We show here that not applying the most accurate filter on the raw data in a much simpler case leads to misleading results. [11] is an example for a problematic "filtering" of the data (the authors actually did not filter the noise in that project). But in not so few projects there are problems with solving the noise, and sometimes also issues with solving the clean data (several examples about this are presented in [29]. See also the discussion in the next part about unique mechanism and the data). Another important issue: very few in this field appreciate the importance of solving the noise while using the properties of the derived clean data. This is one of the concepts in this project:

we build new algorithms that filter the noise while using the likelihood of the obtained clean data.

**I. 2. Solving the noise: Known methods and main difficulties.-** There are many methods that scientists use when solving data from individual molecules [30-58, 61-63]. But existing methods treat mainly trajectories without noise or simple noise forms and binned trajectories. In [38] we showed how we can utilize the information in the clean data (using canonical forms). In this regard, we showed that there are not so few projects with results that are not unique [38]. Here we show: "weak" filters leading to misleading results. And weak filters are not so rare. Here we solve cases where binning the data fails, where binning is used extensively in this field. Our methods filter the noise in the trajectory of the photon times in the order recorded. The filter goes much further than previous methods that were applied on this trajectory [50].

**I. 3. Canonical models.-** There is a problem in this field that was noticed since the 80s [30, 36, 40, 41, 42] and recently solved [37, 38, 39]: a multi substate *on-off* KS is not uniquely obtained from (even a clean & very long) two - state trajectory, and thus we must first construct a canonical form from the data for an accurate analysis (otherwise the result is just one option from many other possibilities that are equivalent statistically). Only one canonical form is built from the data. We have developed very efficient canonical *on - off* mechanisms termed reduced dimensions forms (RDFs) [37, 38, 39]: these are mathematical mechanisms that are (relatively) simply built from clean data yet also from a kinetic scheme. These mechanisms have many advantageous over other methods in accuracy and robustness (these are listed in our papers [37], [38], [39]). Our new filters will use RDFs also in tackling the noise

**I. 4. The idea of the filter.-** The questions are: How can we solve correctly the noise in the data?? How can we extract all the information from the noisy data?? How can we use the information in the noisy data and finding the correct model from the data?? In answering these questions, here we write a filter (a numerical algorithm) based on new numerical and statistical treatments and a new general likelihood function in this field (see codes at, http://www.flomenbom.net/codes_project12.html). We suggest a general new concept: writing the likelihood function in a way that involves the raw data, and the derived clean data, see Eq (3*). In advanced filters, we express the clean data with our recently developed canonical forms, RDFs [37,38,39]. [In fact, we suggest expressing the term involving the correlations among events in the clean data with the appropriate reduced dimensions form, see Eq. (3*)]. We show here that this concept works in solving two state noisy photon trajectories from various kinetic schemes. We find that the specific likelihood form involves a combination of (*a*) the photons (the observable), (*b*) the derived clean data presented with *on* and *off* durations and (*c*) *on off* correlation terms. In most cases, the best results are seen when all these terms in the likelihood function are included in an unbiased way after a special normalization causing all terms having an equal contribution. (In fact the filter is very sensitive to the form of the likelihood functions. The form of the likelihood function determines whether the filter will work rather than fail.) We show that this form of the algorithm is crucial for cleaning accurately the noise (various other variants lead to wrong results). Let me also say and emphasize: in this study we show that the filter must use the clean data while removing the noise, otherwise the results are not accurate.

In the next part of this paper, we first talk about the analysis of the clean data (also presenting RDFs), and then list noise-sources in measurements. In the main part of this chapter, the new filters are presented with results about several dozen systems. In part III we present concluding comments.

## II. Solving noisy trajectories

**II. 1. Analysis of clean trajectories.-** We start with a short description of our toolbox for solving clean trajectories [37,38,39]

### II. 1. 1. The information content in the data.-
Here, we assume that the data is infinite long and without noise. Therefore, we can construct directly from the data any jumping time (JT) probability density function (PDF) (we use in the text either JTPDF or just PDF when talking about other probability density functions). These include: $\phi_x(t)$ and $\phi_{x,y}(t_1, t_2)$, where, $x,y = on, off$. $\phi_x(t)$ gives the probability density that an event is state $x$ lasts time $t$: in other words, $\phi_x(t)$ gives the probability density that the time duration is state $x$ (=$on$, $off$) is $t$. $\phi_{x,y}(t_1, t_2)$ gives the probability density that an event in state $x$ lasts time $t_1$ and the following event in state $y$ lasts time $t_2$. These JTPDFs are expressed with exponential expansions:

$$\phi_x(t) = \sum_i c_{x,i} e^{-\lambda_{x,i} t} \tag{1}$$

and,

$$\phi_{x,y}(t_1, t_2) = \sum_{m,n} \sigma_{x,y,mn} e^{-\lambda_{x,n} t_1 - \lambda_{y,m} t_2}. \tag{2}$$

$\phi_{on}(t)$ and $\phi_{off}(t)$ are constructed from trajectory 1A (Fig. 1A) and are shown in Fig. 1C & 1D, respectively. Eq. (1) is with $L_x$ terms and Eq. (2) with $L_x L_y$ terms. $L_x$ is usually the number of substates in state $x$ in the KS.

### II. 1. 2. Constructing a mechanism from the clean data.-
Assuming $\phi_x(t)$ and $\phi_{x,y}(t_1, t_2)$ are known, we focus on constructing a KS from these JTPDFs. For this, we construct the likelihood function, $l(\Theta)$,

$$l(\Theta) = \sum_{x,y} \sum_i \log\left(\phi_{x,y}(t_{1,i}, t_{2,i})\right), \tag{3}$$

and maximize $l(\Theta)$ with respect to $\Theta$, where $\Theta$ is the set of rates in the KS. In Eq. (3), the index $i$ represents the $i^{th}$ cycle in the actual cleaned *on - off* data. We perform the maximization with constraints: the coefficients in $\boldsymbol{\Theta}$ should also reproduce the coefficients of $\phi_x(t)$. Yet, finding the KS from $\phi_x(t)$ and $\phi_{x,y}(t_1,t_2)$ is difficult. The reasons are: (1) the number of the substates in each of the states, $L_x$ ($x = on,\ off$), is usually large, and (2) the connectivity among the substates is usually complex. Yet, the data has limited information content, and so not *all* the details regarding the KS are obtainable from the data. In addition, there are many local solutions in the landscape of the coefficients [38], many of these solutions are very different than the correct KS. We can average over many initial conditions and this is an exhaustive search since convergence is not guaranteed in the space of coefficients. (3) A fundamental difficulty in finding the correct KS arises from the equivalence of KSs; namely, there are a number of KSs with the same trajectory in a statistical sense [33 - 35, 37, 38, 39].

These three issues form a problem when solving the data.     We solve these issues with canonical (unique) forms [37, 38, 39]. The space of KSs is mapped. The new space in made of canonical forms. A given KS is equivalent with a unique canonical form, yet several KSs can have the same canonical form. KSs with the same canonical form are equivalent, and cannot be discriminated based on the information in (also) an ideal two - state trajectory. We have derived new canonical forms: reduced dimensions forms (RDFs) [37, 38, 39]; see Fig. 1F for an example. RDFs are not Markovian models since the connections in RDFs have multi - exponential JTPDFs. The advantageous of RDFs in solving the problem of relating a model with the time *on - off* trajectory over other approaches are numerous [37,38,39]: RDFs are physical models, accurately constructed from the data, accurately related with a set of KSs, etc.

We can simply construct the RDF from $\phi_x(t)$ and $\phi_{x,y}(t_1,t_2)$. First, we note that the rates in the exponential expansion of the JTPDF $\phi_x(t)$ are the same as those in the exponential expansion of the JTPDFs, $\varphi_{x,mn}(t)$; $\varphi_{x,mn}(t)$ is the JTPDF connecting substates $n_x \rightarrow m_y$ in the RDF, and follows:

$$\varphi_{x,mn}(t) = \Sigma_{\nu=1}^{L_x} \alpha_{x,m\nu n} e^{-\lambda_{x,\nu}t}. \qquad (4)$$

Then, we note that the rank $R_{x,y}$ ($x \neq y$) of the matrices $\sigma_{x,y}$ that appear in the double exponential expansion of $\phi_{x,y}(t_1, t_2)$ gives, in most systems, the number of substates in state $y$ in the RDF. Finally, the coefficients in $\{\alpha_x; \alpha_y\}$ are found when maximizing the likelihood function, Eq. (3), when built from the RDF.

***II. 1. 3. Constructing the RDF from finite length ideal data.-*** Finding the most reliable RDF from a finite trajectory, even without noise, is a real challenge; the reason is that $\phi_x(t)$ and $\phi_{x,y}(t_1, t_2)$ are not known, and we need smart numerical procedures for extracting these PDFs from finite data. We have developed a set of procedures, forming a toolbox, for constructing reliably the RDF from finite data [38,39]. The toolbox executes the following steps: * The rates and the coefficients in the exponential expansion of the JTPDF $\phi_x(t)$s are found with fitting, using a new procedure based on the Padé approximation method. * The ranks $R_{x,y}$s of the matrices $\sigma_{x,y}$s are found from the matrices $\phi_{x,y}(t_1, t_2)$s; any particular $\phi_{x,y}(t_1, t_2)$ has the same rank as of the corresponding $\sigma_{x,y}$, yet, the rank of $\phi_{x,y}(t_1, t_2)$ is much more accurately obtained from finite data. We have developed a new numerical procedure that computes the rank of the $\phi_{x,y}(t_1, t_2)$s from the data. * The matrices $\sigma_{x,y}$ are estimated from the data while constructing special JTPDFs (of the total of following JTs and of the total of the square root of following JTs) with a new numerical procure * The last step uses Eq. (3) when constructed from the RDF.

Using the toolbox, a RDF is constructed from the data fairly accurately, and importantly, much more accurately than other mechanisms. Once the RDF is constructed, we can express this with a set of KSs. The set usually contains the most possible KSs that are associated with the constructed RDF. Choosing from the set a particular KS requires additional information

**II. 2. The noise in the data.-** The problem of dealing with noise is an unsolved issue in the context of data from individual molecules. In this paper, we suggest a filter that solves a particular system (two state photon data and generalizations of any discrete data), yet also a general method that we can use on any data. In this part, we present all the main issues that are related with noise in two-state data and other data - types, where in part **II. 3** we present the filter for solving noisy data.

***II. 2. 1. The type of the external noise.-*** The type of the external noise depends on the measurement's type. This information is used in the analysis of the noisy data. Examples include Poissonian noise and Gaussian noise. In particular: in measurements that collect photons, the time among following photons is monitored. A simple model for generating a photon two - state trajectory is shown in Fig. 2A. The *on - off* Markovian KS has transition rates $\lambda_{on}$ and $\lambda_{off}$, connecting, respectively, the *on* substate with the *off* substate and the *off* substate with the *on* substate. Once the process occupies the *on* (*off*) state it emits photons with a rate, $\gamma_{on}$ ($\gamma_{off}$). In fact, noise photons are recorded also when the process is in the on state

Two - state trajectory can have a Gaussian noise: this is observed e.g. in ion channel recordings. We generate such data while first still generating a *clean* two-state trajectory, $u(t)$, yet here a (zero-averaged with width $\sigma_z$) Gaussian noise $z(t)$ is added every $dt$: the equation for the signal $w(t)$ reads: $w(t) = u(t) + z(t)$.

***II. 2. 2.-*** ***The strength of the noise.-*** Clearly, cleaning correctly the noise depends on the strength of the noise. Indeed, the experimentalists' interest is designing clean measurements. Yet, in the analysis, we must have a way solving any value of the ratio signal/noise, and this includes a way identifying 'too' high noise levels

***II. 2. 3.-*** ***Internal noise: the issue of time resolution, detection efficiency, etc.-*** The noise can also originate from low time - resolution of the experimental devices compared with the measured process. Say that the fastest duration of *on - off* transitions $\tau$ is smaller than the time for detecting the required amount of photons. This means that fast events can be missed. A missed fast *on* event can also originate when the number of photons that are emitted in $\tau$ is the same as the number of noise photons in $\tau$ . A missed *off* event occurs when the number of noise photons emitted in a fast *off* duration $\tau$ is unusually large and thus bridges two *on* events. Small detection efficiency can result in similar problems

***II. 2. 4. Correlations in the noise.-*** The noise may have internal correlations, and also correlations with the state the process staying in. When analyzing the process, we must use the fact that the noise is correlated, that is, for discriminating signal from noise, we must include the existence of correlations in the noise. Not doing this, lead to erroneous results.

***II. 2. 5. Using correlations in the clean data in solving noise.-*** when there are correlations among durations in the clean data (e.g.: $t_{x,i+1}$ is statistically dependent on $t_{x,i}$), we must take in account these correlations when cleaning noisy data. Again, one must use this information for seeing accurate results.

***II. 2. 6.- other issues****.-* each issue that interferes with identifying accurately the state of the process from the observable is noise. For example, entities diffusing in and out the laser spot, fluctuating coefficients (the detection efficiency, etc.), etc. We must treat all these issues

## II. 3. Filtering the noise

In **II. 1 & II. 2**, we have defined the problem: for solving the data in the right way while using RDFs, we need solving the noise in the data. Our basic new mathematical approach presented in part (**I.4**) enable us writing the likelihood function in the way:

$$l(\mathbf{\Theta}) = \Sigma_{m,n} \log(\mathrm{P}_{observ}(m, \mathbf{\Theta}_d)\phi_{clean\ data}(n, \mathbf{\Theta}_c)) + compensation\ terms \qquad (3^*)$$

The symbol Sigma including all observable values and all values of the derived clean data. Thus, the first step in the algorithm is identifying the type of the observable in $P_{observable}(i, \Theta_d)$. Yet, Eq. (3*) determining the best solution (the best identification of the observable), while maximizing $l(\Theta)$, with a fixed $\Theta$. The set $\Theta$ includes the coefficients representing (for example) the photons, $\Theta_d$, and those coefficients representing the derived clean data, $\Theta_c$. We find these coefficients from the data, namely, we write these with statistics from the data. $\phi_{clean\,data}(j, \Theta_c)$ is first approximated with statistical functions. Only in an advanced step in the filtering, $\phi_{clean\,data}(j, \Theta_c)$ representing the model. The model is a RDF. Thus, in this scheme, we iterate among the identification of the photons and the identification of the best mechanism that can generate the clean data. Likelihood functions are frequently appearing with *compensation terms* because there are biases in the likelihood function. Yet, the *compensation terms* in Eq. (3*) are system-dependent and will determine the strength of the filter. In addition, we can perform several adjustments also on the first term in Eq. (3*). In what follows, we show this while solving pretty complicated versions of noisy two-state photon data with the general concepts presented in part **II. 1 & II. 2** resulting in Eq. (3*). The next step is solving higher order discrete data (requiring simple generalizations) and then solving FRET data and continuous data (requiring involved generalizations). We will also create software based on the developed methods

***II. 3. 1. Known statistical methods for noise-filtration.-*** Clearly, many authors dealt with solving the actual data in relevant measurements [46 - 54]. General methods that were suggested: the maximum likelihood technique, a maximum entropy approach, a Bayesian information approach. Quite a few authors dealt with part of the specific problems that were defined in **II. 2** [48 - 54]. For example, correcting for missed fast events using a modified likelihood function was suggested in [54] when tackling ion channel data. Such methods can appear also here, adjusted, when building the RDF from the data.

Yet, let me emphasize that not so few techniques for solving noise in two - state data are partial: simple thresholds and smoothing are usually applied on the binned data. Correlations in the noise, or among the noise and the state of the mechanism are not taken in account. In addition, the methods do not use correlations among following durations in the clean data when solving the noise. In not so few cases, the methods assume that (due to various treatments) the noise is averaged out and does influence the analysis. Clearly, in many cases these assumptions are "too" simple and may lead to misleading conclusions. There are cases where filters that rely on binning fails yet our filter presented here removes the noise (we show such an example in the next pert and in the supplementary file).

Tools for solving noisy $m$ - state trajectories, FRET trajectories, and continuous trajectories are sparser.

***II. 3. 2. Cleaning noisy photon two state data.-*** We present here and in the next part new filters that solve noisy two state photon trajectories: we apply Eq. (3*) on various cases and build the numerical algorithm. This algorithm is containing many innovative treatments.

We say that the measured entity stays at its position while it is measured and that the detection efficiency (quantum yield, etc) is high. Still, the data is rather short and contains just 2700 *on-off* cycles. (Solving cases where the entity diffuses requiring a simple extension in the likelihood function). Indeed, finding solutions for $m$ - *state* trajectories, FRET trajectories, and continuous trajectories is more complicated, nevertheless, since the filters use the basic concepts presented here, these can supply a positive indication about the applicability of our direction in solving also these other systems. This study also shows that primitive filters lead to wrong results. Not applying all steps in this filter (in the way found) may result in rates that are ten times different than the actual rates. This filter tackles issues that were not solved in the past: authors frequently assume that the noise is averaged out, or small, or not important and ignored the noise, or calculate correlations function from the raw data - averaging out most of the new information that is contained in data from individual molecules, and other questionable techniques that lead to misleading or partial results.

We start with the KS in Fig. 2A. We set $\lambda_{on}$=1/10, $\gamma_{on}$=1, $\lambda_{off}$=1/99, $\gamma_{off}$=1/10 (all units are scaled) ($\lambda_x$ is rate jumping from state $x$ and $\gamma_x$ is the photon emission rate in state $x$: see Fig 2A). We generate the clean data and the photon data (Fig 2B, 2C, 2D). This design enabling having 10 photons in any state, in a typical event, and in addition *on* photons are ten times faster, so in most cases there is a clear distinction among states. Nevertheless, in Fig 2D we show that also is such a simple design, there are *on* events that are fast with only one photon (photon 19, and photon 55 in 2D) and when such an event happens, it looks like one long *off* event, rather than: *off* event, short *on*, *off* event. Our filters designed here can solve such cases. Note also that we apply the filter on many sets of these rates finding the interval where results can be accurately seen. This in fact solves the issue of signal to noise ratio and is related with missed fast events in a general way.

The filter is built in the following way[1]:

(1) create the PDF of the photons. Identify the two parts (*on & off*). We must see these parts in this PDF in order having the possibility of filtering the data. The intersection among the parts is a threshold, *TrSld₁*.

(2) Compute directly from the data $1/\gamma_{on}$ and $1/\gamma_{off}$ using *TrSld₁*. Since there are many *off* photons left to *TrSld₁*, we use a special correction formula:

---

[1] ***The codes are presented at*** http://www.flomenbom.net/codes_project12.html

$$\tilde{t}_{off} = \left(1 - e^{-\frac{TrSld_1}{t_{off}}}\right)\left(-TrSld_1 e^{-\frac{TrSld_1}{t_{off}}} + t_{off}\left(1 - e^{-\frac{TrSld_1}{t_{off}}}\right)\right) + t_{off}e^{-\frac{TrSld_1}{t_{off}}}.$$

$\tilde{t}_{off}$ is the updated value and $\tilde{t}_{off} = \frac{1}{\gamma_{off}}$. (In what follows we write $t_{off}$ also the updated quantity)

(3) build a likelihood threshold using the derived rates,

$Tr\_L = 1/(1/t_{on} - 1/t_{off})*\log(t_{off}/t_{on})$.

This is a second threshold, $Tr\_L$. The likelihood threshold is obtained when setting: likelihood *on* photon equal likelihood *off* photon

(4) Build a threshold for a slow *on* photon, $TrOn_3$,

$TrOn_3 = t_{on}*\log(N_p/10)$.

$N_p$ is the total number of photons. $TrOn_3$ is slowest possible *on* photon in the data.

(5) Build trajectories with averaging of order *n*, where *n=2,...,27*. We average any photon with *n* following photons.

(6) Build the photon PDF for each trajectory of order *n* (see Fig 2E). Compute the 3 thresholds in each *n* order trajectory: $TrSld_1$ is still the intersection of the two parts of the photon PDF, $Tr\_L$ is obtained from the *on* part of the photon PDF ($Tr\_L$ is the time that maximizing the *on* part plus one standard deviation) and $TrOn_3$ is obtained from the *off* part of the photon PDF (it is the time that maximizing the *off* part averaged with the intersection of the parts). See Fig 2E.

(7) The algorithm determines each photon in each trajectory of order *n*. It uses the thresholds: following *on* photons are checked with the condition: at least one of the three following photons is smaller than $Tr\_L$ and the actual photon is smaller than $TrOn_3$. Following *off* photons are checked with the condition: at least one of two following photons is larger than $Tr\_L$ or the actual photon is larger than $TrSld_1$.

(8) There is a correction part in the algorithm compensating on the fact that the edges photons (when changing states) are smoother (relative with the actual trajectory) in higher order trajectories

(9) The likelihood function determining the best result. It is a form - sensitive function. We have found the best form here: we first compute the likelihood of the photons in each trajectory, $l_{ph.}(n)$,

$$l_{ph.}(n) = \sum_i \log\left(\gamma_{on} e^{-t_{on,i}(n)\gamma_{on}}\right) + \sum_i \log\left(\gamma_{off} e^{-t_{off,i}(n)\gamma_{off}}\right).$$

Here, $t_{on,i}(n)$ is on photon number $i$ in trajectory of order $n$, and similar in the *off* photons. The exponential function in $l_{photon}(n)$, $\psi_x(t) = \gamma_x e^{-\gamma_x t_x}$, is extendable: the photon emission mechanism, $\psi_x(t)$, can result in, e.g., a multi exponential function, yet, $\psi_x(t)$ can include other effects. We normalize $l_{ph.}(n)$: $l_{ph.}(n) \rightarrow l_{ph.}(n)/\max_n |l_{ph.}(n)|$. We then compute the likelihood of the obtained clean data (this is the part of the *on & off* durations), $l_{data}(n)$,

$$l_{data}(n) = \sum_i \log\left(\lambda_{on} e^{-u_{on,i}(n)\lambda_{on}}\right) + \sum_i \log\left(\lambda_{off} e^{-u_{off,i}(n)\lambda_{off}}\right).$$

Here, $u_{on,i}(n)$ is *on* duration number $i$ in the trajectory with averaging of order $n$, and similar in the *off* durations. Again, the exponential function representing the clean data durations, $\phi_x(t)$, is extendable, and account for any mechanism that generate these times. We normalize $l_{data.}(n)$: $l_{data}(n) \rightarrow l_{data}(n)/\max_n |l_{data}(n)|$. The total likelihood is a simple combination of the two parts. Yet, we also check about correlations in the data and when such are seen, another term, $l_c(n)/\max_n |l_c(n)|$, is included in the total likelihood function (see next part regarding discussion). Thus, the likelihood function in this filter follows:

$$l(\boldsymbol{\Theta}, n) = \tilde{l}_{ph.}\left(\boldsymbol{\Theta}_{ph.}(n), n\right) + \tilde{l}_{data}\left(\boldsymbol{\Theta}_{on\,off}(n), n\right) + \tilde{l}_c(\boldsymbol{\Theta}_c, n) \qquad (5)$$

where the symbol twiddle represents the normalization specified above, and we explicitly write here the dependence of the likelihood function on the coefficients and *n*: the averaging degree.

The new treatments in this filter include: (A) developing the various thresholds (B) creating smooth trajectories of order $n$ (C) developing the conditions in the algorithm (using the thresholds) with the correction-part (D) creating the special likelihood function from Eq. (3*). We emphasize that we can use in steps 6, 7, 8, and 9, other conditions and likelihood functions, namely, this filter is the basic form and we can build many related filters from this one. In fact, we show in what follows, several likelihood functions and several set of conditions in this basic filter, and study the advantages of each of these in cleaning the noise.

***Simple Filter.-*** Applying the filter, we see encouraging results: we are able finding $\lambda_{on}$ & $\lambda_{off}$ in various other cases (see table 1). For example, in the case specified above, the best result finds $\lambda_{on}$ within 23%, $\lambda_{off}$ within 13%, where *on* photons are identified correctly 83% of the time and *off* photons 97% of the time. We also find the region where the data is "too" noisy: in order having accurate results, we must have an average of at least 5 *on* photons and $\gamma_{on}$ is five times faster than $\lambda_{off}$ and $\gamma_{off}$. Otherwise, filtering will fail. In table 1, we also talk about results from three likelihood functions, showing that here Eq. (5) solves the noise, where other forms fail. We also apply a partial filter on several cases showing that we must use the entire filtering. In a "partial" filter, we use a simple threshold - filter with the values, $Tr\_L(n)$ and $TrShld(n)$, instead of the conditioning parts (6, 7, 8). In both cases, the condition checked either the $i$ actual photon and the $i$ averaged photon (e.g., the condition is: photon$(n,i)>= Tr\_L(n)$ | photon$(1,i)>= Tr\_L(1)$, when checking an *off* photon with a photon - likelihood threshold). The results show 33% accuracy in both rates in both cases.

This filter solves cases where binning fails. We show this explicitly in the supplementary file.

***Advanced filter.-*** We apply on the data also an advanced filter, where the conditions in part (7) in the algorithm are changed: (***) here, in the condition part about the *on* photon, we consider three photons. The condition is: either $t_{x,i} < TrSld_1$ or $t_{x,i} < TrOn_3$ with: $t_{x,i+1}(n) < Tr_L(n)$ &

$t_{x,i+2} < TrSld_1$ or $t_{x,i+2}(n) < Tr_L(n)$ & $t_{x,i+1} < TrSld_1$. (***) the *off* photon - condition is about five photons: (A) the *i* photon is larger than *TrSld₁* (B) one in three averaged photons [photon(*n,i*), phtons(*n, i+1*), photon(*n, i+2*)] are larger than *TrSld₁(n)* (C) photon(*n,i+3*) is larger than *TrSld₁(n)* and the average of the previous three photons is larger than *Tr_L(n)* (D) photon(*n,i+4*) is larger than *TrSld₁(n)* and the average of the previous four photons is larger than *Tr_L(n)*.

Table 1 also reporting on the results with the advanced filter. In various cases, the results are better with the advanced filter than the simple filter. This is in cases where the number of *off* photons detected is relative small, yet the difference among the average durations of *on* and *off* photons is a constant and large: $\gamma_{on} / \gamma_{off}$ is large, and, $\gamma_{on} / \lambda_{off} \Rightarrow$ decreasing from a large number towards one. When many photons are detected, yet the difference among the average *on* and *off* photon times is relative small ($\gamma_x / \lambda_x$ is large, and, $\gamma_{on} / \gamma_{off} \Rightarrow$ decrease towards one), the simpler algorithm is better.

***The logic in the basis of the filters***.- we expect seeing many short photons in an *on* state and many slow, long, photons in the *off* state. In principle, a simple cut - off can discriminate *on* and *off* photons in a deterministic photon emission case. Yet, the emission of the photons is also a stochastic process. Namely, in the *on* event we may see several photons that are slow and in an *off* event we may see several fast photons. The second case will occur often when the *off* photon emission PDF is exponential, where the first case is relatively rare. In particular, the probability that an *on* photon is larger than any cut - off *c* is given by: $p_{c,on} = \int_c^\infty \psi_{on}(t)dt$. When, $\psi_{on}(t) = \gamma_{on} e^{-\gamma_{on} t}$, with, e.g., $\gamma_{on} = 1$ and $c = 3.33$, we have, $p_{c,on} = e^{-\gamma_{on} c} \approx 0.033$. Thus here, two consecutive slow photons signal a jump from the *on* state with a probability of 99.9%. In this way, we can choose the conditions in the algorithm. What about the *off* state?? $p_{c,off} = \int_0^c \psi_{off}(t)dt$ is the probability that an *off* photon is faster than *c*. When, $\psi_{off}(t) = \gamma_{off} e^{-\gamma_{off} t}$, with $\gamma_{off} = 1/10$ and $c = 3.33$, we have, $p_{c,off} = 1 - e^{-\gamma_{off}c} \approx 1/3$. Thus, even three fast consecutive *off* photons can appear in the trajectory with a relatively large probability. Just the sixth fast consecutive photons may signal on a jump from the *off* state. Again, we see that the conditions in the algorithm are adjustable depending on $p_{c,off}$. When we study the following fast photons until we see a slow one (in an *off* state), we check the average of these fast photons: the average should exceed (at least) the likelihood threshold (the average of *m* *off* photons is $1/\gamma_{off}$ with a width of $1/(\gamma_{off}\sqrt{m})$, and the average minus the width is much larger the likelihood cut – off also when *m* is 5 (and even smaller) in the case specified above).

We also emphasize that in an advanced filter that we design now, we use also likelihood computations when determining jumps among states.

### II. 3. 3. Cleaning noisy two state photon data with correlations in the cleaned data.-
The data is generated from a KS with four substates: two *on* substates and two *off* states, see Fig 3A. This generates a trajectory with correlations among durations in the clean data. The filter presented in the previous part is applied also here. The only modification includes another term, $l_c(n)$,

$$l_c(n) = \sum_{x,y,i} \log \left( N_{x,y} \exp\left\{ -\left( \sqrt{u_{x,i}(n)u_{y,i'}(n)} - < \sqrt{t_x t_y} > \right)^2 / 2\tilde{\sigma}_{x,y}^2 \right\} \right). \quad (6)$$

Here, $x,y=on,off$, and $i$ is cycle $i$ ($i'=i+1$, and when $x=on$ & $y=off$, $i'=i$), $N_{x,y}$ is a normalization constant, and, $\tilde{\sigma}_{x,y}^2 = < t_x t_y >$. $l_c(n)$ is included in the likelihood function, Eq. (5), and is accounting for correlations in the data. In the likelihood function, we include a normalized term: $l_c(n)/\max_n |l_c(n)|$. This term is included only when the correlation condition shows correlations in the clean data. Thus, we first compute the correlation condition:

$$C_{x,y} = | \frac{<t_x t_y>}{<t_x><t_y>} - 1|.$$

Here, $x$, $y=on$, $off$. Only when $C_{x,y} \geq 0.1$, $\widetilde{l_c}(n)$ is included in the likelihood function. The results are encouraging in the various cases checked (results about the eight cases are presented in table 2). Only when the correlation term is included [in a normalized, unbiased, way, Eq. (5)], we see accurate results. The likelihood function is presented in Fig 3B. At the solution that maximizes the likelihood function, we find: (1) 99% *on* photon identification and 86% *off* photons, (2) the average duration in the *on* state is found within 24%, and 45% in the *off* state (3) Correlations coefficients in durations (derived from the cleaned data) are found within 67%. Positive correlation condition is seen.

In Table 2, we present all additional results about this system. One important conclusion is that when the correlation signal is small (that is, smaller than 0.1, and there are cases here that this is also the mathematical value), a different normalization than that in Eq. (5) can work better (that is, we find better results with another normalization): this operation normalizing each term with the number of events (for example, the *on* photon − likelihood is normalized with the number of *on* photon detected, etc) rather than with the largest absolute value (depending on *n*) of the term. This operation increases the influence of correlation term, and can help increasing the accuracy of the filter in correlated data with a relative small correlation condition.

### III Concluding remarks

***Ground breaking nature of the work.-*** noise in the data is not treated accurately in not so few projects reported in the literature in this field every year. Here we show: not solving the noise accurately results in misleading conclusions. We also show that there are cases where binning the data fails (with any noise filtration method that relied on binning) yet the filter presented here working. Solving the noise in relevant data will help significantly many. (A) The groundbreaking mathematical achievements here are the development of a new form of the likelihood function, Eq. (3*), and its specific forms in discrete data. We showed here that in two state noisy photon trajectories, this form is a combination of three terms: the photon data, the clean *on off* durations, and the correlations. Each term is normalized in a special way: in most cases, the normalization follows: $l_*(n)/\max_n |l_*(n)|$. There are cases where we normalize with the specific number of events per term. We see in preliminary studies of a different filter that the likelihood function should also contain specially designed compensating terms in order having accurate results (B) The uniqueness in the statistical and technical fronts is the algorithms' forms containing several new treatments. (C) the new mathematical methods and numerical algorithms developed here combined with our previous results in this field, will enable us create software. This is an important part in this project continuation. Further related filters will appear in future publications.

***Further research.-*** The filter introduced here cleans the noise. Once we have the clean data, we can use our toolbox of finding the model from clean data [38] [see also the part (II. 1)]. Even before, we can apply another filter that uses the results from the clean data of the first filter. We are about finalizing such a filter. This filter is used on correlated data. It uses the data, the averaged data, the three thresholds, plus another threshold: the upper bound when two following *on* photons are slower than *TrSld$_1$*. The idea here is using the various thresholds while determining the photon's type. This filter first builds groups of following *on* photons and *off* photons depending on *TrSld$_1$*. Then, the type of each group is updated depending on a set of conditions involving the thresholds and the local types of photon groups. There are also cases where we must compute the local likelihood of three following groups (we check all eight (*on* - *off*) possibilities), and choose the case with the maximal likelihood. This filter is classier since it uses explicitly the correlations in the data and its design is more complicated (the filter uses local information on several groups when determining the type of the group and thus is based on correlations). We can apply this filter in an iterative way (the thresholds are updated after the filter is applied, and the filter is used again until convergence is seen). Tackling other trajectory – types will demand further development.

We also plan collaborating with experimentalists when tackling the noise with the new filters: we will solve trajectories that measure the dynamics and activity of three glycosyltransferases [60] with experiments involving single molecules techniques (AFM, FRET and other spectroscopic methods). This collaboration will help creating stable filters.

**References.-**

## Figure caption

**FIG 1** A segment of a clean two-state trajectory (**A**), the noisy data (**B**), $\phi_{on}(t)$ (**C**), $\phi_{off}(t)$ (**D**), a KS (**E**) and its corresponding RDF (**F**). Trajectory **A** is generated when simulating a random walk in KS **1E**, and adding a zero mean Gaussian noise for every element of the clean data, with a variance, $\sigma_z$=37/100 counts/bin. $\phi_{on}(t)$ and $\phi_{off}(t)$ are shown, on a log-linear scale, in panels **C** & **D**, respectively. The KS **1E** has $L_{on}$=2 (squared substates), $L_{off}$=2 (circled substates), and irreversible transitions. For using the KS (e.g. for generating the data, and for fully constructing the RDF from it) we assign numerical values for the connections among substates in the KS (the connections are termed reaction rates or transition rates or kinetic rates). The RDF is obtained from the KS with a clear - defined mathematical mapping developed in [37].

**FIG 2** A KS with a mechanism of photon emission (**A**), the clean *on-off* trajectory (**B**), the binned data (**C**), the photon trajectory (**D**), and log linear plot of the steady state histogram of photons (**E**). In **A**, a curly arrow indicates on the emission of photons and a full arrow is a transition rate. The rates obey: $\lambda_{on}$=1/10 and $\lambda_{off}$=1/99, $\gamma_{on}$=1 and $\gamma_{off}$=1/10 (scaled units). The binned trajectory in **C**

is very difficult to work with. The photons in **D** are plotted in the order detected, where *on* (*off*) photons are in red (green). The algorithm is applied on this trajectory. The histogram in **E** is the photon PDF in order $n$=8. Its two parts are clear

**FIG 3** The data was generated from the KS **A** with the following rates ($\lambda_{x,21}$ connecting substates 1 in $x$ and 2): $\lambda_{off,21}$ =0.85/99, $\lambda_{off,11}$=0.15/99, $\lambda_{off,12}$=0.85/499, $\lambda_{off,22}$=0.15/499, and, $\lambda_{on,21}$ =0.85/9, $\lambda_{on,11}$=0.15/9, $\lambda_{on,12}$=0.85/99, $\lambda_{on,22}$=0.15/99. Once the process is in any of the *off* substates, it emits photons with a rate, $\gamma_{off}$=1/10, where when staying in the *on* substates, photons are emitted with a rate, $\gamma_{on}$=1. The likelihood function is presented in B, on a log linear scale. Here, the maximal likelihood value is when $n$=12

# Figure 1

**A**

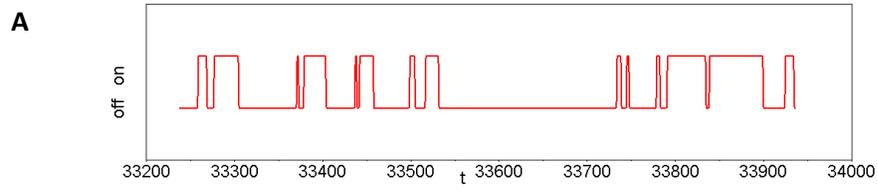

**B**

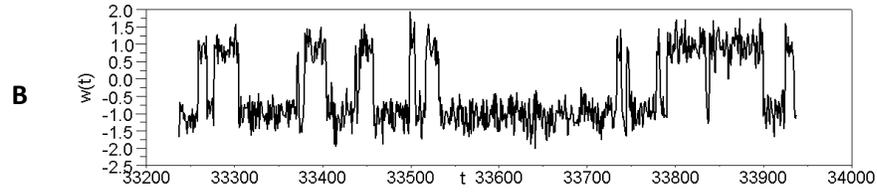

**C**

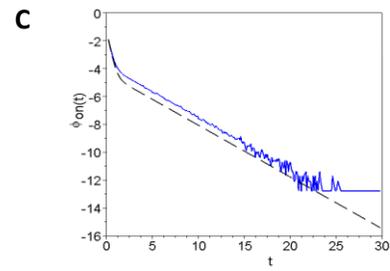

**D**

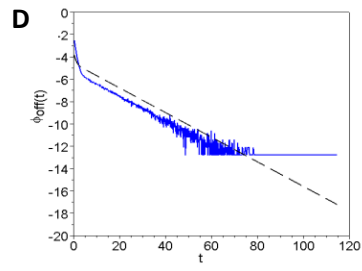

**E**

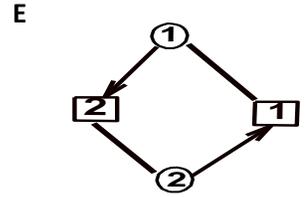

**F**

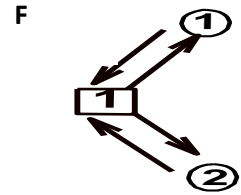

**Figure 2**

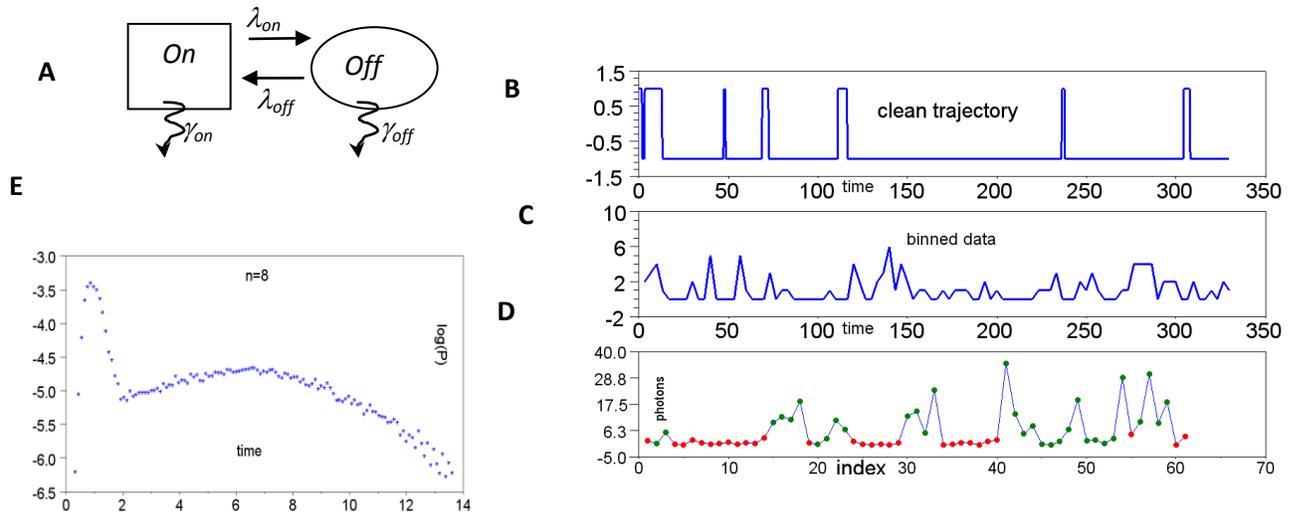

A

$\lambda_{on}$

On    Off

$\lambda_{off}$

$\gamma_{on}$    $\gamma_{off}$

B    clean trajectory

C    binned data

D    photons

E    n=8

log(P)

time

time

index

**Figure 3**

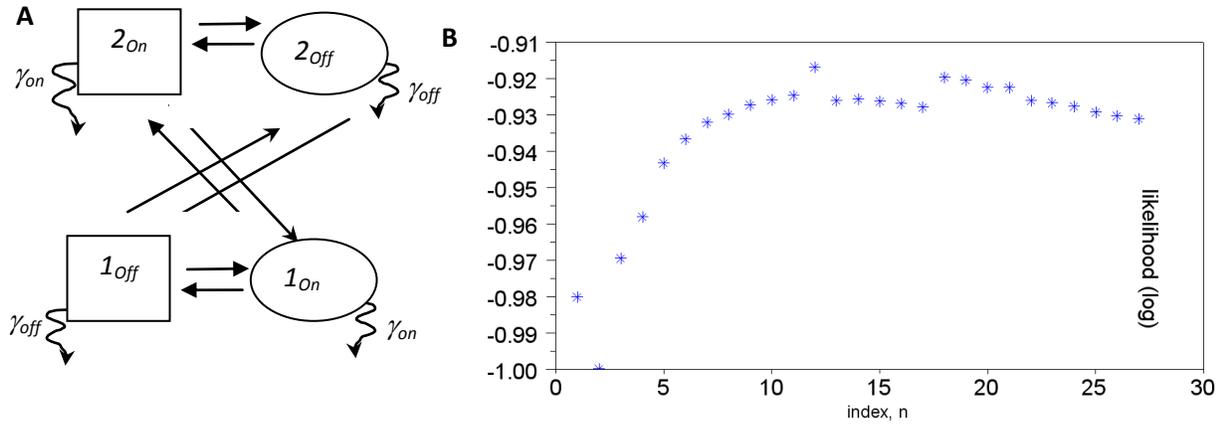

# Table 1

| Case | Likelihood 3 |
|---|---|
| *(1)* $\lambda_{on}$=1/10, $\gamma_{on}$=1, $\lambda_{off}$=1/99, $\gamma_{off}$=1/10 | $p(\lambda_{on})$=108%, $p(\lambda_{off})$=93%, 94% *off*, 86% *on* |
| *Advanced filter on case 1* | $p(\lambda_{on})$=98%, $p(\lambda_{off})$=88%, 94% *off*, 85% *on* |
| *(2)* $\lambda_{on}$=1/10, $\gamma_{on}$=1, $\lambda_{off}$=1/33, $\gamma_{off}$=1/10 | $p(\lambda_{on})$=139%, $p(\lambda_{off})$=193%, 92% *off*, 91% *on* |
| *Advanced filter on case 2* | $p(\lambda_{on})$=123%, $p(\lambda_{off})$=103%, 91% *off*, 86% *on* |
| *(3)* $\lambda_{on}$=1/10, $\gamma_{on}$=1, $\lambda_{off}$=1/5, $\gamma_{off}$=1/10 | $p(\lambda_{on})$=416%, $p(\lambda_{off})$=163%, 74% *off*, 96% *on* |
| *Advanced filter on case 3* | $p(\lambda_{on})$=198%, $p(\lambda_{off})$=123%, 72% *off*, 89% *on* |
| *(4)* $\lambda_{on}$=1/49, $\gamma_{on}$=1, $\lambda_{off}$=1/49, $\gamma_{off}$=1/10 | $p(\lambda_{on})$=113%, $p(\lambda_{off})$=103, 93% *off*, 98% *on* |
| *Advanced filter on case 4* | $p(\lambda_{on})$=87%, $p(\lambda_{off})$=116%, 98% *off*, 96% *on* |
| *(5)* $\lambda_{on}$=1/49, $\gamma_{on}$=1, $\lambda_{off}$=1/49, $\gamma_{off}$=1/3.77 | $p(\lambda_{on})$=103%, $p(\lambda_{off})$=63, 92% *off*, 86% *on* |
| *Advanced filter on case 5* | $p(\lambda_{on})$=113%, $p(\lambda_{off})$=42%, 98% *off*, 84% *on* |
| *(6)* $\lambda_{on}$=1/10, $\gamma_{on}$=1, $\lambda_{off}$=1/60, $\gamma_{off}$=1/10 | $p(\lambda_{on})$=126%, $p(\lambda_{off})$=103, 99% *off*, 97% *on* |
| *Advanced filter on case 6* | $p(\lambda_{on})$=157%, $p(\lambda_{off})$=94, 98% *off*, 92% *on* |
| *(7)* $\lambda_{on}$=1/6.1, $\gamma_{on}$=1, $\lambda_{off}$=1/60, $\gamma_{off}$=1/10 | $p(\lambda_{on})$=134%, $p(\lambda_{off})$=106, 95% *off*, 84% *on* |
| *Advanced filter on case 7* | $p(\lambda_{on})$=164%, $p(\lambda_{off})$=119, 94% *off*, 78% *on* |
| *(8)* $\lambda_{on}$=1/2.77, $\gamma_{on}$=1, $\lambda_{off}$=1/60, $\gamma_{off}$=1/10 | $p(\lambda_{on})$=207%, $p(\lambda_{off})$=123, 94% *off*, 75% *on* |
| *Advanced filter on case 8* | $p(\lambda_{on})$=233%, $p(\lambda_{off})$=179%, 89% *off*, 72% *on* |

**Table 1** The results when filtering the data from the simple mechanism in Fig 2A: $\lambda_{on}$ & $\lambda_{off}$ in terms of the real values, and the percentage of correct determination of the type of the photon. We checked the results with three likelihood function. The results are from likelihood function 3 (unless otherwise is indicated). **Likelihood 1**: the normalization of the function is performed with the number of cycles. The results are like in case 3, since here the main contribution in the likelihood function is from the photon - likelihood. **Likelihood 2**: every term in the likelihood function is normalized with the number of events in the specific $\Sigma$. Here, in most cases the likelihood in an increasing function of $n$ resulting in relatively large rates. There are cases where we see a maximal value at an intermediate $n$ resulting from the patterns of the detection probability: the *off* detection probability is usually a decreasing function of $n$, where the *on* detection probability is an increasing function of $n$. Yet, there are cases where this likelihood function is better than other functions, and this is when the correlation signal is small. **Likelihood 3**: normalizing every term in $\Sigma$ with the largest value in all $n$, see Eq. (5). The results are reported in the table

# Table 2

| Case | Likelihood 3 |
|------|--------------|
| $\lambda_{on,1}$=1/9, $\lambda_{on,2}$=1/99, $\lambda_{off,1}$=1/99, $\lambda_{off,2}$=1/499 | $P(t_{on})$=124%, $p(t_{off})$=145%, 86% *off*, 99%*on*, $Pc$=*[69,71]%*, $SC$=*[9,9,6,1]%* |
| *Advanced filter on case 1* | $P(t_{on})$=135%, $p(t_{off})$=134%, 98% *off*, 96%*on*, $Pc$=*[64,78]%*, $SC$=*[4,25,3,1]%* |
| *(2)*$\lambda_{on,1}$=1/19, $\lambda_{on,2}$=1/81, $\lambda_{off,1}$=1/49, $\lambda_{off,2}$=1/199 | $P(t_{on})$=108%, $p(t_{off})$=132%, 74% *off*, 99%*on*, $Pc$=*[78,78]%*, $SC$=*[8,9,8,4]%* |
| *Advanced filter on case 2* | $P(t_{on})$=121%, $p(t_{off})$=117%, 93% *off*, 98%*on*, $Pc$=*[80,98]%*, $SC$=*[10,14,12,5]%* |
| *(3)*$\lambda_{on,1}$=1/5, $\lambda_{on,2}$=1/19, $\lambda_{off,1}$=1/49, $\lambda_{off,2}$=1/99 | $P(t_{on})$=147%, $p(t_{off})$=249%, 72% *off*, 99%*on*, $Pc$=*[33,33]%*, $SC$=*[1,0,1.9,2.3]%* |
| *Advanced filter on case 3* | $P(t_{on})$=183%, $p(t_{off})$=203%, 83% *off*, 93%*on*, $Pc$=*[29,30]%*, $SC$=*[2,3,0,2.7]%* |
| *(4)*$\lambda_{on,1}$=1/5, $\lambda_{on,2}$=1/5, $\lambda_{off,1}$=1/49, $\lambda_{off,2}$=1/199 | $P(t_{on})$=167%, $p(t_{off})$=450%, 72% *off*, 97%*on*, $Pc$=*[13,14]%*, $SC$=*[0,0.6,1,2]%* |
| *Advanced filter on case 4 (likelihood 2)* | $P(t_{on})$=184%, $p(t_{off})$=119%, 95% *off*, 75%*on*, $Pc$=*[45,45]%*, $SC$=*[1,1,2,5]%* |
| *(5)*$\lambda_{on,1}$=1/9, $\lambda_{on,2}$=1/81, $\lambda_{off,1}$=1/49, $\lambda_{off,2}$=1/199 | $P(t_{on})$=99%, $p(t_{off})$=116%, 99% *off*, 69%*on*, $Pc$=*[97,98]%*, $SC$=*[3,3,1,0]%* |
| *Advanced filter on case 5* | $P(t_{on})$=125%, $p(t_{off})$=127%, 98% *off*, 98%*on*, $Pc$=*[69,68]%*, $SC$=*[1,6,5,0]%* |
| *(6)*$\lambda_{on,1}$=1/19, $\lambda_{on,2}$=1/81, $\lambda_{off,1}$=1/49, $\lambda_{off,2}$=1/199 | $P(t_{on})$=111%, $p(t_{off})$=133%, 99% *off*, 69%*on*, $Pc$=*[69,69]%*, $SC$=*[3,2,1,2]%* |
| *Advanced filter on case 6* | $P(t_{on})$=119%, $p(t_{off})$=116%, 98% *off*, 96%*on*, $Pc$=*[77,76]%*, $SC$=*[3,3,4,0.67]%* |
| *(7)* $\lambda_{on,1}$=1/5, $\lambda_{on,2}$=1/19, $\lambda_{off,1}$=1/49, $\lambda_{off,2}$=1/99 | $P(t_{on})$=184%, $p(t_{off})$=135%, 97% *off*, 99%*on*, $Pc$=*[42,42]%*, $SC$=*[0,0,0,1.5]%* |
| *Advanced filter on case 7* | $P(t_{on})$=167%, $p(t_{off})$=174%, 87% *off*, 94%*on*, $Pc$=*[36,35]%*, $SC$=*[0,3.8,1,2.61]%* |
| *(8)* $\lambda_{on,1}$=1/5, $\lambda_{on,2}$=1/5, $\lambda_{off,1}$=1/49, $\lambda_{off,2}$=1/199 | $P(t_{on})$=600%, $p(t_{off})$=136%, 67% *off*, 99%*on*, $Pc$=*[13,12]%*, $SC$=*[1.4,3,4.9,5]%* |
| *Advanced filter on case 8 (likelihood 2)* | $P(t_{on})$=144%,$p(t_{off})$=123%, 96% *off*, 81%*on*, $Pc$=*[58,57]%*, $SC$=*[1.4,1.8,1,1.3]%* |

Table 2 Results about filtering the trajectories from Kinetic Scheme 3A. In all cases: $\gamma_{on}$=1, $\gamma_{off}$=1/10. In the first 4 cases, the jumping probabilities from substates of the same states are 15%, where in cases 5 until 8 these probabilities equal 33%. Namely, in cases 1 until 4, the probability of jumping from substate $1_{on}$ and reaching substate $1_{off}$ is 85%, where the jumping times are exponentially distributed with a rate $\lambda_{on,1}$. *** Likelihoods are similar with those presented in the previous table. *** Here, $P(t_{on})$ is the computed average *on* durations in terms of the mathematical value. *** The term $Pc$ containing the averages <$t_{on}, t_{off}$> and <$t_{off}, t_{on}$> in terms of the mathematical values. *** $SC$ are the correlation conditions in all combinations of $x,y$=*on,off* (*on on, on off, off on, off off*). *** We see that when the correlation signal is low, likelihood 2 is better in various cases

## Supplementary information:

## Solving single molecules: filtering noisy discrete data made of photons and other type of observables

*Ophir Flomenbom, Flomenbom-BPS Ltd, 19 Louis Marshal St., Tel Aviv, Israel 66268*

In the supplementary file, I show: the filter presented in the main text solves cases where binning the data doesn't work: the obtained binned trajectory is "too" noisy and all the methods that are used on the binned data will not help. (The methods in [50] do not filter the noise: these calculate correlation functions from the raw data (and from slightly smoothed data) but this does not clean the data and the extracted information is rather poor in content).

In the figures we simply show that when the noise increases there are cases where binning results in trajectories where an *off* event might contain a peak that is identical with *on* events and that there are *on* events that contain "too" few photons, identical with an *off* event. Thus any method that is applied on the binned data will not filter such cases.

We highlight the point that here the noise is since also in the *off* state photons are recorded with an exponential rate. Fourier transformed the binned data will not help much here, in particular in the cases specified in the next figures.

Just an algorithm that study every photon and using the local information of consecutive photons and global information (thresholds extracted from the entire data) can help. The algorithm of the main text constituting the best way that enabling solving the data photon after photon.

***A simple case.***

In figure 1 in the supplementary material, we present data that is generated from the mechanism in figure 2A in the main text. The rates are: $\lambda_{on}$=1/10,$\gamma_{on}$=1, $\lambda_{off}$=1/99, $\gamma_{off}$=1/10 (all units scaled). This figure is with various panels: the upper panel is the clean data and the binned data: number of photons in a bin of size 1.99. The other (lower) panel is the photon durations in the order recorded and we also plot the real identification: the larger value is the *about $1/\gamma_{off}$* and the smaller value is the about *$1/\gamma_{on}$*

This set up generating very clear two state data: we see the two states also in the binned data (middle of the figure, blue) and the binned data & the clean data (red curve) coincide. Thus cleaning the noise from the binned data is simple. We show with arrows events that are missed (green: missed *on* event, black: missed *off* event) when thresholding the binned data with threshold = 3.

The lower panel showing the raw data, photon durations in the order recorded (blue) with the real identification (green). Clearly this trajectory is smoother than the binned data. Although in this case also the binned data is not "too" noisy we show here cases where filtering the raw data is simple but the binned data is too noisy: these are labeled with # and @ in both panels.

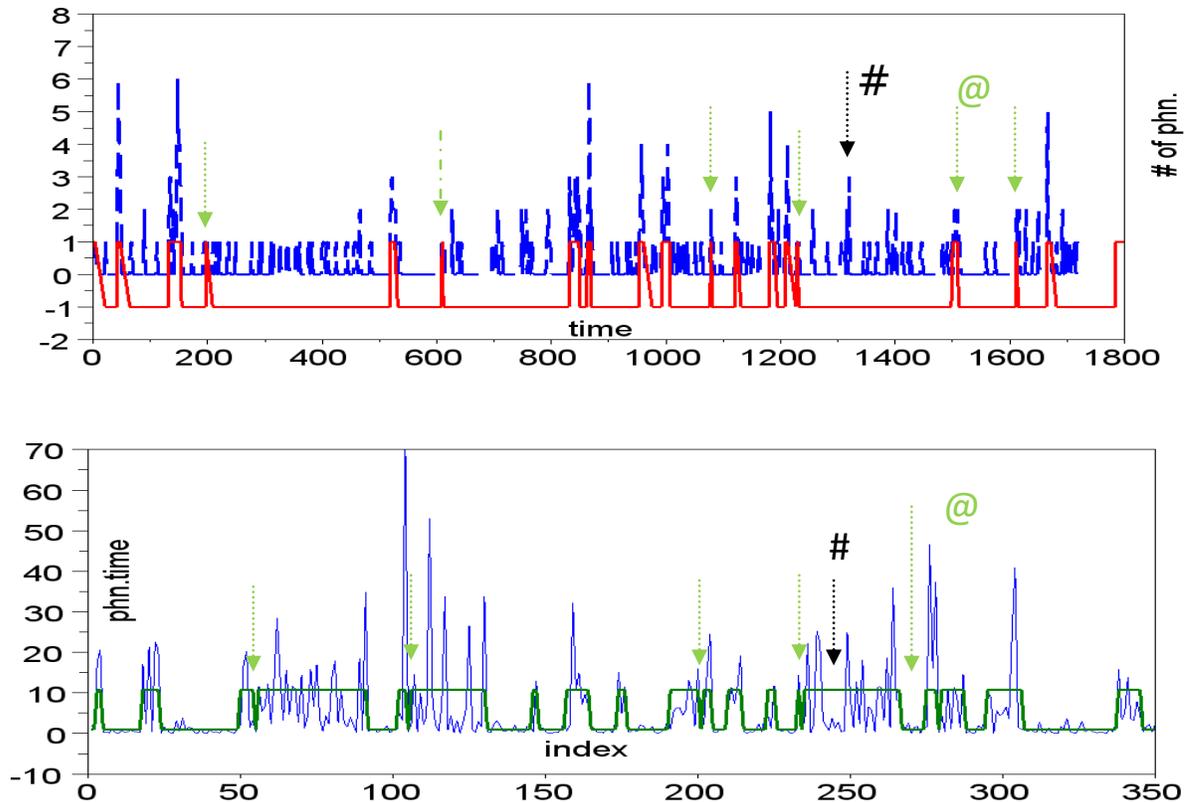

***Figure 1 supplementary information.-*** The binned data (upper panel), and the raw data (lower panel), the photon durations. In both panel, we show also the clean data. Description is presented in the text

***A more complicated case.***

The data was generated from KS 2A, yet the rates are: $\lambda_{on}$=1/10,$\gamma_{on}$=1, $\lambda_{off}$=1/49, $\gamma_{off}$=1/10. I have changed the rate controlling the *off* durations: here this rate is just 1/49 and in the previous example $\lambda_{off}$ is 1/99. This resulting in many *off* events with few photons. This increases the noise: there are not so few *off* events that do not have the required amount of photons needed in order seeing a very slow one. In general, we see here that the binned data is rather noisy, clearly relative with the raw data representation of photon durations in the order recorded. I solved such data with my methods (see table 1). In the figure, we show with arrows all the cases where the binned data is *on* but smaller than the threshold: threshold =3. We show these cases also in the raw data representation. At least case 2 &3 (chronological order) are solved with the filter presented in the main text yet are missed in the binned data with any filtering technique. This is at least 25% better.

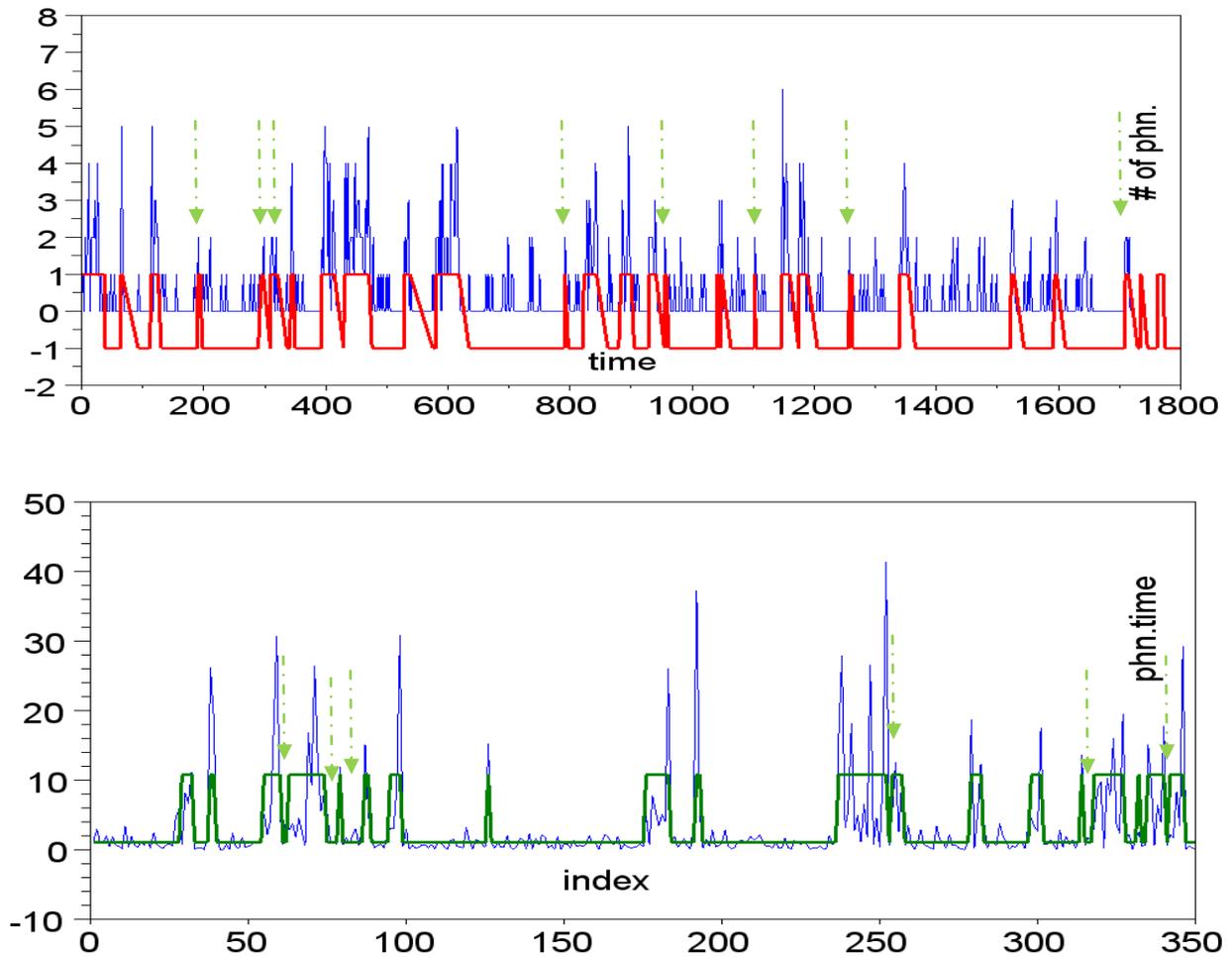

***Figure 2 supplementary information.-*** The description is presented in the text

***The complicated case.***

The data was generated from KS 2A, yet the rates are: $\lambda_{on}$=1/10, $\gamma_{on}$=1, $\lambda_{off}$=1/33, $\gamma_{off}$=1/10. I have changed the rate controlling the *off* durations: here this rate is just 1/33 and in the previous example $\lambda_{off}$ is 1/49.

This set up resulting in many *off* events with few photons: shorter events than in the previous example. This increases the noise: there are not so few *off* events that do not have the required amount of photons needed in order seeing a very slow one. Again: in general, we see here that the binned is rather noisy, clearly relative with the raw data representation of photon durations in the order recorded. I solved also such data with my methods (see table 1). In the figure, we show with arrows all the cases where the binned data is *on* but smaller than the threshold: threshold =3. These are shown in green. We show with a black arrow when having an *off* event with many photons in a bin: larger than 3. We show all these cases also in the raw data representation. At least case 1 & 2 & 9 (chronological) are solved with the filter presented in the main text yet are missed in the binned data with any filtering technique. This is at least 33% better. We show in panel 3 that also increasing the bin size does not improve the situation: one additional *on* event is identified but one *off* event has many photons in a bin.

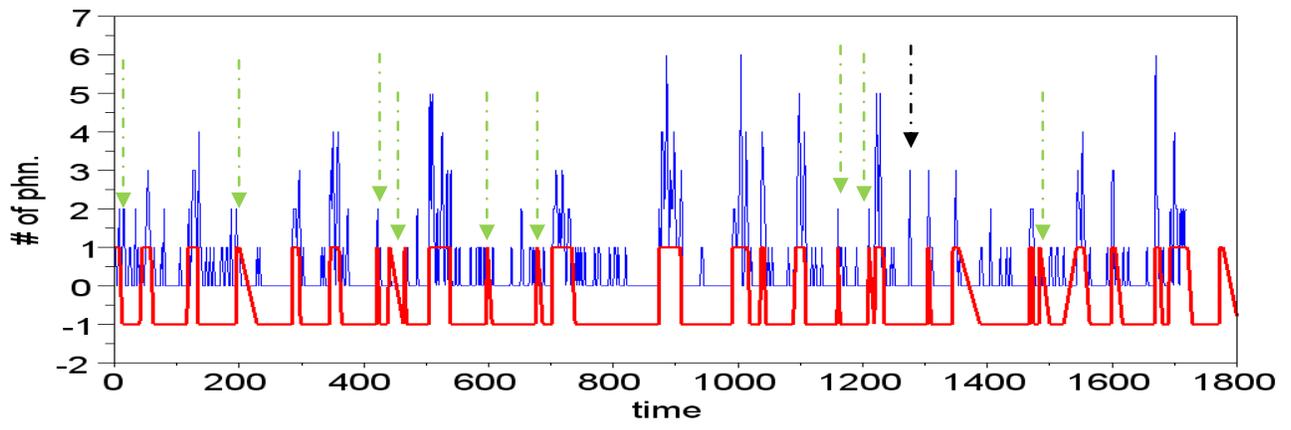

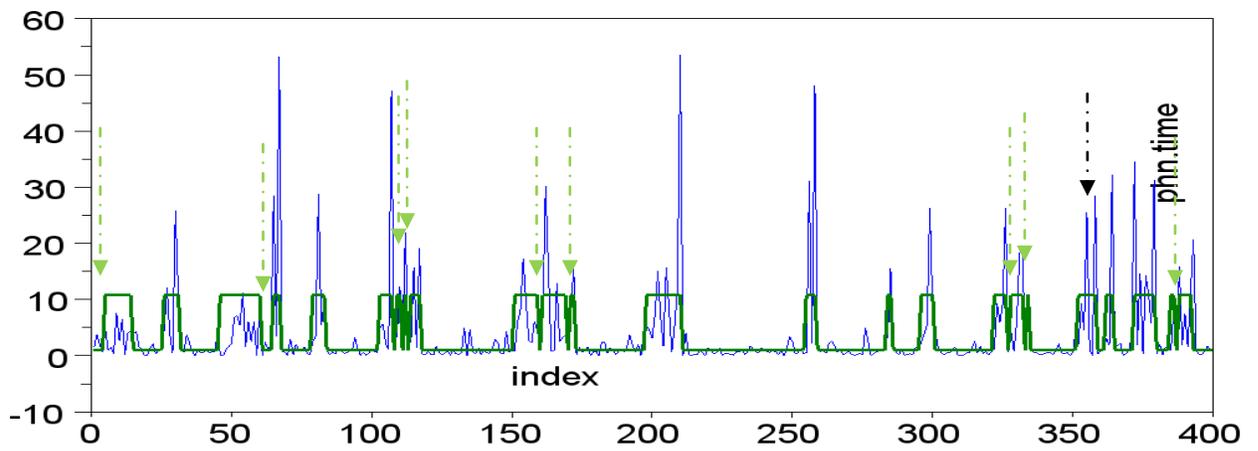

Larger bin: bin = 3.99

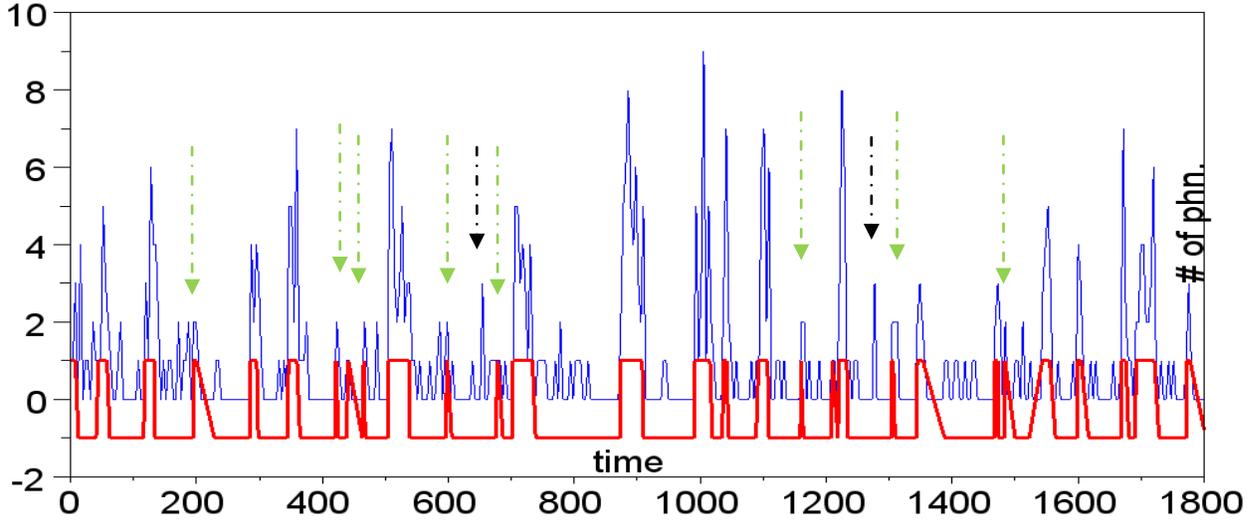

*Figure 3 supplementary information.-* The description is presented in the text